\documentclass[10pt,leqno]{amsart}
\usepackage{graphicx}
\usepackage{indentfirst,csquotes}

\usepackage{amssymb,amsthm,amsmath}
\usepackage{xcolor,paralist,hyperref,titlesec,fancyhdr,etoolbox}
\usepackage{secdot}
\usepackage{booktabs}
\usepackage[numbers]{natbib}

\titleformat{\section}[display]{\normalfont\huge\bfseries\centering}{\centering}{10pt}{\Large}
\titlespacing*{\section}{0pt}{0ex}{0ex}

\hypersetup{ colorlinks=true, linkcolor=black, filecolor=black, urlcolor=black }

\usepackage{lipsum}
\newcommand{\fnm}[1]{#1}
\begin{document}
\title{Attention-Enhanced Reinforcement Learning for Dynamic Portfolio Optimization} %%%%%%%%%%%%
\author[1]{\fnm{Pei Xue} }\email{peix324@yahoo.com}

\author[2]{\fnm{Yuanchun Ye} }\email{yuanchuy@andrew.cmu.edu}

\date{Sep, 2025}
\address{Tepper School of Business, Carnegie Mellon University, Pittsburgh, PA, USA}

\begin{abstract}
We propose a deep-reinforcement learning framework for dynamic portfolio optimization that combines a Dirichlet policy with cross-sectional attention mechanisms. Unlike softmax- or projection-based approaches, the Dirichlet distribution enforces feasibility by construction, accommodates tradability masks, and provides a coherent geometry for exploration. Our architecture integrates per-asset temporal encoders with a global attention layer, allowing the policy to adaptively weight sectoral co-movements, factor spillovers, and other cross-asset dependencies. Economic realism is incorporated through a reward function that penalizes turnover-based transaction costs and portfolio variance, directly linking the agent’s objective to classical mean–variance trade-offs.
We evaluate the framework on a comprehensive S\&P 500 panel from 2000 to 2025 using purged walk-forward backtesting to prevent look-ahead bias. Empirical results show that attention-enhanced Dirichlet policies deliver higher terminal wealth, Sharpe and Sortino ratios than equal-weight and reinforcement learning baselines, while maintaining realistic turnover and drawdown profiles. Our findings highlight that principled action parameterization and attention-based representation learning materially improve both the stability and interpretability of reinforcement learning methods for portfolio allocation.
\end{abstract} %%%%%%%%%
\maketitle

\bigskip

\section{Introduction}

Designing robust portfolio allocation under uncertainty is a central problem in financial economics. Classical frameworks, beginning with mean–variance optimization \cite{markowitz1952portfolio} and intertemporal extensions \cite{merton1971optimum}, prescribe allocations by trading off expected return against risk. Although analytically elegant, these models rely on restrictive assumptions about return distributions, stationarity, and preferences, and typically neglect trading frictions. Subsequent work sharpened risk measurement (e.g., CVaR, drawdown) and improved covariance estimation via shrinkage, factor structures, and robust optimization. Yet even with these advances, most approaches remain essentially myopic: they solve a one-period problem repeatedly rather than address sequential decisions as market conditions evolve.

Reinforcement learning (RL) provides a natural way to cast portfolio choice as a Markov decision process (MDP). The investor observes a state summarizing recent market history, chooses an action (portfolio weights), and earns a reward reflecting portfolio performance net of frictions. Policy-gradient and actor–critic methods directly optimize long-run performance. A growing literature applies RL to trading and allocation with encouraging results \cite{moody2001learning,jiang2017deep,yang2020deep,zhang2020finrl}, but many studies use ad-hoc action parameterizations (e.g., softmax with post-hoc projection), omit realistic frictions, or evaluate in limited settings.

We revisit portfolio RL with a focus on principled action parameterization and rigorous economic evaluation. We propose a \emph{flat Dirichlet policy} that models portfolio weights directly on the simplex, enforcing the budget and non-negativity constraints by construction, handling tradability masks naturally, and supplying a coherent geometry for exploration. We embed economically motivated frictions in the reward: (i) per-period turnover costs to capture transaction costs, and (ii) a quadratic variance penalty $w_t^\top \Sigma_t w_t$ to discourage concentration in correlated assets. This links the RL objective to classical portfolio trade-offs while retaining sequential adaptivity. Our evaluation uses purged, walk-forward backtesting (to avoid look-ahead), reports \emph{after-cost} Sharpe/Sortino, drawdown, and turnover, and runs factor regressions to test whether performance reflects genuine alpha rather than rediscovered exposures.

By combining a principled policy distribution, realistic cost and risk modeling, and stringent evaluation, we clarify what RL contributes to portfolio management once ad-hoc design choices are stripped away. While we focus here on a \emph{non-hierarchical} (flat) action structure to isolate these ingredients, the framework is extensible: richer state encoders (e.g., temporal transformers, cross-asset attention) and hierarchical actions can be layered on the same foundation. Even in the flat setting, enforcing feasibility at the action level and embedding economic frictions materially improves stability and economic realism relative to standard RL baselines.

\section{Related Literature}

\subsection{Classical Portfolio Theory and Dynamic Extensions}
The modern theory of portfolio choice originates with mean–variance optimization \cite{markowitz1952portfolio}, which trades expected return against variance under quadratic preferences or normal returns. Continuous-time formulations generalize to intertemporal settings with stochastic opportunities and CRRA preferences \cite{merton1971optimum}. Kelly-style log-utility and growth-optimal strategies \cite{kelly1956new} emphasize long-run geometric growth and motivate log-return objectives akin to our reward.

Portfolio risk has advanced along two axes: \emph{measurement} and \emph{regularization}. On measurement, variance remains tractable but is tail-insensitive; coherent, tail-sensitive measures—especially CVaR—admit convex formulations and are widely used \cite{rockafellar2000optimization}. Path-dependent preferences motivate drawdown constraints, which control the time profile of losses rather than cross-sectional dispersion \cite{grossman1993optimal}. On regularization, high-dimensional estimation makes sample covariances ill-conditioned; shrinkage toward structured targets and factor-based models improve conditioning and out-of-sample stability \cite{ledoit2004honey}. Robust-optimization variants hedge against moment uncertainty. Within this tradition, quadratic penalties $w^\top \Sigma w$ remain reliable and interface cleanly with modern covariance regularization, while CVaR/drawdown often serve as complements in robustness checks.

Trading frictions fundamentally shape optimal policies. Dynamic models with linear and nonlinear costs (temporary and permanent impact) show that rebalancing schedules and turnover constraints are as critical as mean–variance trade-offs \cite{almgren2000optimal,gatheral2010no}. Our reward therefore includes per-period turnover costs so the agent internalizes implementation shortfall.

\subsection{Predictability, Factors, and Cross-Sectional Signals}
Expected returns exhibit persistent cross-sectional patterns (value, size, momentum, profitability, investment), typically summarized by factor models such as three- and five-factor specifications \cite{fama1993common,fama2015five}. Time-series and cross-sectional momentum \cite{jegadeesh1993returns} and volatility clustering \cite{bollerslev1986generalized} are well documented. The proliferation of proposed factors has raised data-mining concerns \cite{harvey2016and}. In practice, portfolio construction fuses multiple signals, balances factor exposures, and controls risk via covariance regularization. Our state includes rolling cross-sectional features and market covariates to capture such structure and its regime dependence.

\subsection{From Control to Learning: MDPs and RL for Portfolio Management}
Casting allocation as an MDP enables sequential optimization under evolving states, risk, and costs. Early work studied recurrent RL for trading \cite{moody2001learning} and order execution \cite{nevmyvaka2006reinforcement}. With deep networks, policy-gradient and actor–critic methods have been explored for allocation and trading \cite{deng2016deep,jiang2017deep,yang2020deep,zhang2020finrl}. We align with policy-gradient RL—notably PPO \cite{schulman2017proximal}—but depart in two ways: (i) the action is a \emph{distribution on the simplex} (Dirichlet), ensuring feasibility and principled exploration; (ii) the reward explicitly incorporates transaction costs and a covariance-based risk penalty, tying learning to standard economic trade-offs.

\subsection{Action Parameterizations for Simplex-Constrained Decisions}
Common implementations map network outputs through softmax and then project to enforce masks or caps, coupling exploration across coordinates and introducing projection-induced discontinuities. By contrast, a Dirichlet policy places mass directly on the simplex; concentration parameters control dispersion and thus exploration in a geometrically appropriate way, with analytic log-probabilities and reparameterized sampling for policy-gradient updates. We adopt a \emph{flat} Dirichlet policy in this first study to isolate action feasibility and exploration geometry.

\subsection{Representation Learning: Sequence Models and Attention}
Financial series display long memory, regime shifts, and cross-asset comovement. Deep sequence models (CNNs, RNNs/LSTMs) and attention-based architectures (e.g., temporal fusion transformers, long-horizon variants) have proven effective in multivariate time series \cite{lim2021temporal,zhou2021informer}. In quant finance, attention and graph neural networks capture cross-asset structure (sectors, supply chains). Our state tensor $X_{t-W+1:t}\in\mathbb{R}^{W\times N\times F}$ plus market covariates and tradability masks is compatible with such encoders; here we keep the policy head flat (Dirichlet) to cleanly assess its contribution.

\subsection{Risk, Costs, and Constraints in Learning-Based Allocation}
Practical allocation must internalize frictions (turnover penalties, no-trade bands) and constraints (no shorting, position/leverage caps); impact models inform cost calibration \cite{almgren2000optimal}. Encoding constraints at the \emph{action} level reduces training pathologies. Our design enforces budget and non-negativity by construction, masks untradable assets, optionally applies per-name caps via capped-simplex projection, and regularizes correlated concentration through $w^\top \Sigma w$.

\subsection{Backtesting Pitfalls and Evaluation Protocols}
Backtest overfitting and data-snooping are pervasive \cite{bailey2014backtest,bailey2016deflated,harvey2016and}. Robust evaluation requires (i) realistic frictions; (ii) walk-forward splits with embargo; (iii) \emph{after-cost} reporting of Sharpe/Sortino, maximum drawdown, and turnover; (iv) factor regressions for attribution; and (v) multiple-testing corrections via reality-check or SPA tests \cite{white2000reality,romano2005stepwise}. We adopt a purged walk-forward design with costs in-reward and report factor-adjusted performance to test for genuine alpha.

\subsection{Positioning of Our Contribution}
Our work sits at the intersection of (a) dynamic portfolio choice with risk and costs, (b) RL for sequential allocation, and (c) representation learning for multivariate financial time series. Relative to prior portfolio RL studies relying on softmax heads or projection, we contribute a \emph{flat Dirichlet policy} that (i) respects the simplex and masks by construction, (ii) provides a coherent exploration geometry, and (iii) integrates with an economically motivated reward (log growth net of transaction costs and a covariance penalty). Focusing on the non-hierarchical setting clarifies the incremental value of these ingredients under rigorous evaluation.

\section{Data and Preprocessing}

\subsection{Source and Coverage}
We use a daily panel of U.S. equities drawn from the S\&P~500 universe. The sample spans approximately two decades (2000--2025). To mitigate survivorship bias, we use historical index constituents rather than the most recent list. Corporate actions (splits, dividends) are adjusted in the raw price series.

\subsection{Data Schema}
The raw dataset is stored in CSV format with the following key columns:
\[
\{\texttt{Date},\; \texttt{ticker},\; \texttt{Open},\; \texttt{High},\; \texttt{Low},\; \texttt{Close},\; \texttt{Volume},\; \ldots \}.
\]
Dates are parsed as \texttt{datetime}, and the panel is sorted by $\texttt{Date}$ and $\texttt{ticker}$.

\subsection{Panel Construction}

The raw CSV file is in \emph{long format}, where each row corresponds to a single stock on a single trading day. To use the data in our model, we convert this into a \emph{panel format} where all stocks for a given day are aligned side by side. 

\begin{itemize}
    \item \textbf{Close prices.} We create a wide matrix of adjusted close prices with dimensions $T \times N$, where $T$ is the number of trading days and $N$ is the number of tickers. Each row corresponds to a calendar date and each column to a stock. 

    \item \textbf{Log-returns.} We compute log-return columns directly from close prices:
    \[
    \text{LR}_{t,i} = \log \frac{\text{Close}_{t,i}}{\text{Close}_{t-1,i}},
    \]
    where $\text{Close}_{t,i}$ is the closing price of stock $i$ on day $t$. We then transform log-returns into simple percentage returns as
    \[
    r_{t,i} = e^{\text{LR}_{t,i}} - 1.
    \]
\end{itemize}

\paragraph{Stacking features.}
We repeat this procedure for each feature column (e.g., volume, RSI, moving averages). Each feature produces a $T \times N$ matrix. These matrices are then \emph{stacked together} along a new dimension to form a three-dimensional tensor
\[
X \in \mathbb{R}^{T \times N \times F},
\]
where $F$ is the total number of features. Thus, for each trading day $t$, $X_{t,:,:}$ contains an $N \times F$ matrix describing all features for all stocks on that day.

\subsection{Cross-Sectional Standardization}
To stabilize learning, we apply cross-sectional z-scoring within each day:
\[
Z_{t,i,f} = \frac{X_{t,i,f} - \mu_{t,f}}{\sigma_{t,f} + 10^{-8}},
\]
where $\mu_{t,f}$ and $\sigma_{t,f}$ are the mean and standard deviation of feature $f$ across all tradable assets on day $t$. Missing values are replaced with zeros.

\subsection{Tradability Mask}
We construct a binary mask
\[
M_{t,i} = \mathbb{I}\{\text{Close}_{t,i}\ \text{is finite}\},
\]
so that untradable assets are excluded. In the environment, weights on masked assets are forced to zero and reallocated across the feasible set.

\subsection{State Windows}
For reinforcement learning, the state at day $t$ is a rolling window of length $W$:
\[
s_t = \{ Z_{t-W+1:t},\; M_t,\; m_t \},
\]
with $Z_{t-W+1:t} \in \mathbb{R}^{W \times N \times F}$, mask $M_t \in \{0,1\}^N$, and optional market-level covariates $m_t \in \mathbb{R}^K$ (e.g., VIX, factor returns).

\subsection{Calendar Alignment}
All panels are reindexed on a common trading calendar and ticker universe. Non-trading days for an asset result in $M_{t,i}=0$, rather than imputation, ensuring that allocations into unavailable assets are prohibited.

\subsection{Data Splits}
We adopt a walk-forward protocol: 2000--2016 is used for training, 2017--2019 for validation and hyperparameter tuning, and 2020--2025 for out-of-sample testing. This chronological walk-forward split enables us to prevent look-ahead bias.

% Table~\ref{tab:data_summary} summarizes the dataset used in the experiments.

\begin{table}[h!]
\centering
\caption{Summary of the S\&P~500 daily panel dataset.}
\label{tab:data_summary}
\begin{tabular}{lcc}
\hline
 & Value & Notes \\
\hline
Time span & 2000-01-04--2025-08-20 & daily data \\
Number of trading days ($T$) & 6446 & After calendar alignment \\
Number of unique tickers ($N$) & 482 & All historical constituents \\
Average tradable names per day & 389.5 & Entry/exit accounted \\
Number of features ($F$) & 19 & Prices, volume, indicators \\
State window length ($W$) & 30 & e.g., 30 days \\
\hline
\end{tabular}
\end{table}

The preprocessing pipeline produces a standardized tensor $Z \in \mathbb{R}^{T \times N \times F}$, a rolling state window of length $W$, and a tradability mask $M_t$ for each day. This construction faithfully mirrors the implementation: pivoting into panels, log-return computation, cross-sectional standardization, mask-aware handling of missing data, and alignment across the trading calendar.
\section{Economic Formulation and Environment}

\subsection{Portfolio Choice as a Sequential Decision Problem}
We cast the daily portfolio allocation problem as a finite-horizon Markov Decision Process (MDP). At each trading day $t$, the agent observes a \emph{state} $s_t$, selects an \emph{action} $a_t$ corresponding to portfolio weights, and receives a \emph{reward} $R_t$ reflecting realized returns net of transaction costs and risk penalties. This formulation aligns with classical portfolio theory \citep{markowitz1952portfolio} but extends it by embedding the allocation problem in a reinforcement learning framework.

\subsection{State Space}

The definition of the state is critical because it determines how much information the agent can condition on when making portfolio choices. We represent the state at time $t$ as a structured object that aggregates both temporal and cross-sectional information over a rolling window of length $W$:
\[
s_t = \{ X_{t-W+1:t}, \; M_t, \; m_t \},
\]
where:
\begin{itemize}
    \item $X_{t-W+1:t} \in \mathbb{R}^{W \times N \times F}$ is a three-dimensional tensor capturing the last $W$ days of observations for $N$ assets across $F$ features. Each slice $X_{\tau}$ for $\tau \in [t-W+1,t]$ provides a full cross-section of the market at a given day.
    \item $M_t \in \{0,1\}^N$ is a tradability mask indicating which assets can be traded at time $t$. This accounts for real-world issues such as missing data, delisted firms, suspended stocks, or corporate events. When $M_{t,i}=0$, the agent is prevented from allocating weight to asset $i$ at time $t$.
    \item $m_t \in \mathbb{R}^K$ are market-level covariates or factors that summarize global conditions not specific to any one stock. These variables allow the policy to incorporate systemic information that often drives cross-sectional returns.
\end{itemize}

This state construction has several advantages. First, the inclusion of a $W$-day rolling window allows the agent to model temporal dependencies such as momentum, reversal, and volatility clustering. Second, the explicit cross-sectional dimension ($N$) enables the policy to exploit contemporaneous relationships across assets, including comovement and lead--lag effects. Third, the tradability mask ensures feasibility under real-world market frictions, while market-level covariates provide a bridge between firm-specific signals and macroeconomic dynamics. 

Compared to static one-period inputs often used in benchmarks, our state representation is richer and closer to the economic decision problem: investors typically condition on both recent time-series history and cross-sectional relative value signals when rebalancing portfolios. This design ensures that the Markov property is approximated: conditional on $(X_{t-W+1:t}, M_t, m_t)$, the distribution of next-period returns is assumed independent of earlier history, making the environment amenable to reinforcement learning algorithms.

\subsection{Action Space}

In our formulation, the action at each decision date $t$ corresponds to the choice of a feasible portfolio vector. The agent outputs daily portfolio weights
\[
a_t = w_t = (w_{t,0}, w_{t,1}, \dots, w_{t,N}) \in \Delta^{N+1},
\]
where $\Delta^{N+1}$ denotes the $(N{+}1)$-dimensional unit simplex. Here $w_{t,0}$ represents the weight on cash (a risk-free position), and $w_{t,i}$ for $i=1,\dots,N$ are allocations to the risky assets in the universe.

The action space is shaped by a set of economically motivated feasibility conditions:
\begin{itemize}
    \item \textbf{Budget constraint:} $\sum_{i=0}^N w_{t,i} = 1$.
    \item \textbf{Non-negativity:} $w_{t,i} \geq 0$, which rules out short-selling and ensures interpretability of allocations as shares of wealth.
    \item \textbf{Cash asset:} including $w_{t,0}$ allows the agent to modulate exposure to risky assets in volatile regimes, akin to dynamic risk budgeting.
    \item \textbf{Tradability:} assets with $M_{t,i}=0$ must satisfy $w_{t,i}=0$, guaranteeing feasibility under suspensions, delistings, or missing data.
    \item \textbf{(Optional) position caps:} per-name caps $w_{t,i} \le \bar{c}_i$ can be enforced to control concentration.\footnote{When used, we enforce caps by projecting the proposed weights onto the capped simplex.}
\end{itemize}

To satisfy the simplex constraints by construction, we model actions as draws from a \emph{flat} Dirichlet distribution:
\[
w_t \sim \mathrm{Dirichlet}(\alpha_0,\alpha_1,\dots,\alpha_N),
\]
which directly places a probability distribution on $\Delta^{N+1}$. Compared to projecting unconstrained vectors back onto the simplex, the Dirichlet parameterization ensures actions are always valid, avoids discontinuities due to projection, and simplifies the handling of transaction costs and tradability masks. The concentration parameters control dispersion across names and thus provide a natural exploration mechanism for policy-gradient methods.

\subsection{Reward Function}

The reward function specifies the objective that the reinforcement learning agent seeks to maximize. In our setting, it captures both the realized profitability of the portfolio and the economic frictions associated with implementing trading strategies. We define the instantaneous reward at time $t$ as
\[
R_t = \log\!\big(1 + w_t^\top r_t \big) \;-\; \kappa \, \text{TC}_t \;-\; \lambda \, w_t^\top \Sigma_t w_t,
\]
where the three terms respectively represent (i) portfolio growth, (ii) transaction costs from rebalancing, and (iii) optional risk penalties. Each component is motivated below.

\paragraph{Portfolio log-return.}
The first term, $\log(1 + w_t^\top r_t)$, is the log-utility of the gross portfolio return. Here $r_t \in \mathbb{R}^N$ denotes the vector of simple one-period returns for the $N$ risky assets, and $w_t$ are the portfolio weights (including cash). The log specification is common in financial economics because it is additive across time, rewards compounding growth, and implicitly encodes risk aversion. Maximizing the sum of log-returns is equivalent to maximizing long-run geometric growth of wealth.

\paragraph{Transaction costs.}
The second term penalizes portfolio turnover:
\[
\text{TC}_t = \sum_{i=1}^N \big|w_{t,i} - \tilde{w}_{t,i}^{\text{pre}}\big|,
\]
where $\tilde{w}_{t}^{\text{pre}}$ are the drifted portfolio weights before rebalancing. This term measures the absolute trading volume required to reach the new target weights. The hyperparameter $\kappa > 0$ represents the effective cost per unit turnover, incorporating bid--ask spreads, brokerage fees, and market impact. By penalizing turnover, the agent learns to balance the benefit of adjusting the portfolio against the cost of trading.

\paragraph{Risk penalty.}
The final term introduces a quadratic risk adjustment, $w_t^\top \Sigma_t w_t$, where $\Sigma_t$ is the sample covariance matrix of returns estimated over a rolling window of length $L$. This expression measures the ex-ante portfolio variance. The penalty weight $\lambda \geq 0$ controls the strength of risk aversion. Economically, this term parallels the variance penalty in mean--variance optimization and discourages the agent from concentrating allocations in highly correlated assets.

\paragraph{Dynamic perspective.}
The reward is evaluated every day, and the reinforcement learning agent maximizes the expected discounted sum of rewards over the full horizon. This transforms the static trade-off of mean--variance optimization into a sequential decision problem: the agent must consider not only the immediate reward but also the impact of current actions on future costs, risk, and state dynamics. This dynamic perspective is particularly valuable in markets where transaction costs, volatility, and correlations evolve over time.

\subsection{Transition Dynamics}

In reinforcement learning, the \emph{transition dynamics} describe how the environment moves from the current state $s_t$ to the next state $s_{t+1}$ after the agent chooses an action $a_t = w_t$. In the portfolio context, these dynamics are governed jointly by realized market returns and the bookkeeping of portfolio weights.

At each trading day $t$, the following steps occur:
\begin{enumerate}
    \item \textbf{Portfolio update.} The agent selects portfolio weights $w_t$. Assets then realize one-period returns $r_t \in \mathbb{R}^N$. The portfolio grows accordingly, and drifted pre-trade weights $\tilde{w}_{t+1}^{\text{pre}}$ are computed to capture changes in relative asset values before rebalancing at $t{+}1$.
    \item \textbf{Feature window roll.} The market feature tensor is updated by discarding the oldest observation and appending the newest one, producing $X_{t-W+2:t+1}$. This rolling window provides the temporal context for the next decision.
    \item \textbf{Tradability mask update.} The mask $M_{t+1}$ is refreshed to reflect which assets are available to trade at the next step. Delistings, suspensions, or missing data cause entries of $M_{t+1}$ to flip to zero.
    \item \textbf{Market-level covariates.} Aggregate signals such as factor returns, volatility indices, or macro variables are recomputed, forming the new vector $m_{t+1}$.
\end{enumerate}

Thus, the next state is
\[
s_{t+1} = \{ X_{t-W+2:t+1}, \; M_{t+1}, \; m_{t+1} \}.
\]

Strictly speaking, asset returns are not fully Markovian: tomorrow’s distribution may depend on long histories, latent factors, or structural breaks. However, by constructing the state as a sufficiently rich rolling window of features, we approximate the Markov property:
\[
P(s_{t+1} \mid s_t, a_t) \approx P(s_{t+1} \mid s_1,\dots,s_t, a_1,\dots,a_t).
\]
In other words, conditional on $(X_{t-W+1:t}, M_t, m_t)$, the distribution of $s_{t+1}$ depends only on the current state and action, not on the entire past. This approximation is standard in financial reinforcement learning and is sufficient for policy-gradient methods such as PPO.

This transition structure captures two essential aspects of real-world portfolio management. First, portfolio weights naturally drift over time as asset prices evolve, creating path-dependence and turnover. Second, the state representation links decisions to evolving market conditions: by rolling windows, tradability masks, and systemic factors, the agent always bases decisions on a realistic information set. The Markov approximation therefore allows the problem to be modeled as a tractable MDP while retaining the key economic features of sequential investment under uncertainty.

\section{Method and Policy Architecture}
\label{sec:method-flat}

\subsection{Overview}
This section specifies a stochastic policy for daily portfolio choice that maps observed market state to a \emph{distribution} over portfolio weights on the simplex. The architecture has three blocks:
(i) a \textit{per-asset temporal encoder} that turns a $W{\times}F$ window into an embedding,
(ii) a \textit{cross-sectional mixer} that models relations across assets at date $t$, and
(iii) a \textit{single Dirichlet action head} over \([\,\text{cash} + \text{all assets}\,]\).

\subsection{Setting and Notation}
At date $t$ there are $N_t$ tradeable equities. For each asset $i$ we observe a $W{\times}F$ window of features
\[
X_{i,t-W+1:t} \;=\; [\,x_{i,t-W+1},\ldots,x_{i,t}\,]\in\mathbb{R}^{W\times F},
\]
a tradability mask $m_t\in\{0,1\}^{N_t}$ (1 if tradeable), and optional market factors $z_t\in\mathbb{R}^{K}$. The state is
\[
s_t \;=\; \big( X_{1,t-W+1:t},\ldots,X_{N_t,t-W+1:t},\; m_t,\; z_t \big).
\]
Actions are portfolio weights $w_t\in\mathbb{R}^{N_t+1}_{\ge 0}$ (index $0$ denotes cash), subject to
\[
\sum_{i=0}^{N_t} w_{t,i}=1,\qquad w_{t,i}=0\ \text{ if }\ m_{t,i}=0\ \ (i\ge 1).
\]
The policy is a conditional distribution $\pi_\theta(\cdot\mid s_t)$ over such $w_t$.

\subsection{Encoders: Temporal (time \texorpdfstring{$\to$}{→} token) and Cross-Sectional (assets \texorpdfstring{$\leftrightarrow$}{↔} assets)}
\paragraph{Temporal encoder (per asset).}
Project features to model width $d$ with $W_{\mathrm{in}}\in\mathbb{R}^{F\times d}$:
\[
u_{i,\tau} \;=\; x_{i,\tau}\, W_{\mathrm{in}} \in \mathbb{R}^{d},\qquad \tau=t\!-\!W\!+\!1,\ldots,t.
\]
\textit{Option A: LSTM.} Apply an LSTM along time and take last/mean hidden state, then project to $d$:
\[
(h^{(\ell)}_{i,\tau},c^{(\ell)}_{i,\tau})=\mathrm{LSTM}_\theta(u_{i,\tau},h^{(\ell)}_{i,\tau-1},c^{(\ell)}_{i,\tau-1}),\quad
h_{i,t}=\tilde h_{i,t}\,W_{\mathrm{out}}+b_{\mathrm{out}}\in\mathbb{R}^{d}.
\]
\textit{Option B: Temporal Transformer.} Add sinusoidal positions $P\in\mathbb{R}^{W\times d}$ and apply $L_{\mathrm{time}}$ encoder layers with multi-head self-attention over time (mask missing rows if needed):
\[
Z^{(0)}_{i,\tau}=u_{i,\tau}+P_\tau,\quad
Z^{(\ell+1)}_{i}=\mathrm{LN}\!\big(Z^{(\ell)}_{i}+\mathrm{MHSA}(Z^{(\ell)}_{i})\big),\quad
Z^{(\ell+1)}_{i}=\mathrm{LN}\!\big(Z^{(\ell+1)}_{i}+\mathrm{FFN}(Z^{(\ell+1)}_{i})\big),
\]
then pool across $W$ (last or mean) to get $h_{i,t}\in\mathbb{R}^{d}$. We share encoder weights across assets and stack
\[
H_t \;=\; [\,h_{1,t},\ldots,h_{N_t,t}\,]^\top \in \mathbb{R}^{N_t\times d}.
\]

\paragraph{Cross-sectional mixer (permutation-invariant).}
To capture co-movement and relative strength across assets on date $t$, we apply a Transformer encoder \emph{across} tokens. Prepend a learnable global token $g_0\in\mathbb{R}^{d}$ and form
\[
T^{(0)}_t \;=\; [\,g_0;\; h_{1,t};\ldots;h_{N_t,t}\,]\in\mathbb{R}^{(1+N_t)\times d}.
\]
Apply $L_{\times}$ encoder layers with self-attention across tokens (no positional encodings on assets to preserve permutation invariance), masking untradable asset tokens:
\[
\begin{aligned}
Q=T^{(\ell)}_t W_Q,\; K=T^{(\ell)}_t W_K,\; V=T^{(\ell)}_t W_V,\qquad
A=\mathrm{softmax}\!\big(\tfrac{QK^\top}{\sqrt{d_k}}+M_{\times}\big),\ U=A\,V,\\
\tilde T^{(\ell)}_t=\mathrm{LN}(T^{(\ell)}_t+U W_O),\quad
T^{(\ell+1)}_t=\mathrm{LN}(\tilde T^{(\ell)}_t+\mathrm{FFN}(\tilde T^{(\ell)}_t)).
\end{aligned}
\]
Let $T^\star_t=T^{(L_{\times})}_t$ and extract the \emph{global summary} $g^\star_t=T^\star_t[0]\in\mathbb{R}^{d}$ and \emph{asset summaries} $A^\star_t=T^\star_t[1:]\in\mathbb{R}^{N_t\times d}$. Optionally inject macro factors via $g^\star_t \leftarrow g^\star_t + W_{\mathrm{mkt}} z_t$ with $W_{\mathrm{mkt}}\in\mathbb{R}^{K\times d}$.

\subsection{Action Head: Single Dirichlet over \texorpdfstring{[cash + all assets]}{}}
From $(g^\star_t, A^\star_t)$ the actor produces logits for cash and each asset:
\[
\ell^{(\mathrm{cash})}_t = a_{\mathrm{cash}}^\top g^\star_t + b_{\mathrm{cash}}\in\mathbb{R},\qquad
\ell^{(\mathrm{asset})}_{t,i} = a_{\mathrm{asset}}^\top A^\star_{t,i} + b_{\mathrm{asset}}\in\mathbb{R},\ i=1,\ldots,N_t.
\]
Stack and map to Dirichlet concentrations (with small $\varepsilon>0$ for stability):
\[
\alpha^{(\mathrm{flat})}_t \;=\; \mathrm{softplus}\!\Big(\,[\,\ell^{(\mathrm{cash})}_t,\ \ell^{(\mathrm{asset})}_{t,1},\ldots,\ell^{(\mathrm{asset})}_{t,N_t}\,]\,\Big) + \varepsilon \;\in\; \mathbb{R}^{N_t+1}_{>0}.
\]
Sample during training (or use the mean at evaluation):
\[
p^{(\mathrm{flat})}_t \sim \mathrm{Dirichlet}\!\big(\alpha^{(\mathrm{flat})}_t\big),
\qquad \mathbb{E}\big[p^{(\mathrm{flat})}_t\big]=\alpha^{(\mathrm{flat})}_t/(\mathbf{1}^\top \alpha^{(\mathrm{flat})}_t).
\]
Apply the tradability mask ($\tilde m_t=(1,m_t^\top)^\top$; cash always feasible) and renormalize:
\[
w_t \;=\; \frac{p^{(\mathrm{flat})}_t \odot \tilde m_t}{\mathbf{1}^\top \big(p^{(\mathrm{flat})}_t \odot \tilde m_t\big)} \;\in\; \Delta^{N_t}.
\]

\subsection{Critic and PPO Training}
A value head $V_\phi(s_t)$ is computed from $g^\star_t$. With advantages $\hat A_t$ (GAE-$\lambda$) and old log-probs $\log\pi_{\theta_{\mathrm{old}}}(a_t\!\mid s_t)$, PPO maximizes
\[
\max_{\theta}\ \ \mathbb{E}\!\left[
\min\!\Big(\rho_t(\theta)\,\hat A_t,\ \mathrm{clip}(\rho_t(\theta),1-\epsilon,1+\epsilon)\,\hat A_t\Big)
\right]
-\eta\,\mathrm{KL}\big(\pi_\theta\|\pi_{\theta_{\mathrm{old}}}\big),
\quad \rho_t(\theta)=\exp\!\big(\log\pi_\theta(a_t|s_t)-\log\pi_{\theta_{\mathrm{old}}}(a_t|s_t)\big),
\]
with value loss $\tfrac{1}{2}(V_\phi(s_t)-\hat V_t)^2$ and optional Dirichlet entropy. The policy log-density is
\[
\log\pi_\theta(a_t\mid s_t) \;=\; \log \mathrm{Dirichlet}\!\big(p^{(\mathrm{flat})}_t;\, \alpha^{(\mathrm{flat})}_t\big),
\]
evaluated at the (sampled) pre-mask $p^{(\mathrm{flat})}_t$ used to form $w_t$.

\subsection{Masking, Constraints, and Frictions}
Budget and non-negativity hold by construction; tradability is enforced via $\tilde m_t$ and renormalization. Optional name caps are enforced by projecting $w_t$ onto a capped simplex. Transaction costs enter the per-period reward through turnover; optional risk penalties use a rolling covariance estimated from recent returns.

\subsection{Complexity and Engineering Notes}
Temporal LSTM scales as $\mathcal{O}(B\,N_t\,W\,d)$ per layer; temporal Transformer as $\mathcal{O}(B\,N_t(W\,d^2 + W^2 d))$. Cross-sectional attention scales as $\mathcal{O}(B\,(N_t{+}1)^2 d)$. We use (i) a learned input projection $W_{\mathrm{in}}$ so $F$ can change, (ii) asset-wise chunking in the temporal encoder, and (iii) masking to keep memory bounded when $N_t$ is large.

\paragraph{Ablations.}
We report variants with (a) LSTM vs temporal Transformer for $f_\theta$, (b) removing cross-sectional attention (temporal-only), and (c) different $W$ and cost levels. This isolates the effect of temporal memory and cross-asset interaction in a \emph{single-pool} setting (no hierarchy).

\section{Experimental Design}\label{sec6}

\subsection{Data, Universe, and Features}
Our empirical study uses daily U.S.\ equity data from the S\&P~500 universe, obtained via the Yahoo Finance public API.\footnote{Accessed at \url{https://finance.yahoo.com}.} For each trading day $t$ and ticker $n$, the panel records adjusted open, high, low, close, and volume (OHLCV), alongside a set of widely used technical indicators. These include 30- and 60-day moving averages, oscillators such as RSI and CCI, directional movement measures $+DI$, $-DI$, and ADX, momentum/trend indicators such as MACD (signal and histogram), and Bollinger bands. The raw files are cleaned by coercing dates into \texttt{datetime}, removing invalid rows, and sorting by $(\text{Date},\text{ticker})$. Close-to-close returns
\[
r_{t+1,n} = \frac{P_{t+1,n}}{P_{t,n}} - 1
\]
serve as price relatives and the basis for reward calculations. Delistings and suspensions are tracked via a tradability mask that prevents allocation to unavailable assets.

Each feature $f$ is pivoted to a $\text{Date}\times\text{ticker}$ matrix and reindexed on the full date–ticker grid. Stacking across features yields a tensor $X\in\mathbb{R}^{T\times N\times F}$ with mask $M\in\{0,1\}^{T\times N}$. To align feature scales while preserving cross-sectional signals, we apply same-day $z$-scoring within each feature:
\[
Z_{t,:,f} = \frac{X_{t,:,f}-\mu_{t,f}}{\sigma_{t,f}+\varepsilon}, \quad \varepsilon=10^{-8},
\]
with NaNs and infinities replaced by zero. This preserves relative momentum/strength information while suppressing level effects.

The sample is split chronologically: letting $t_{\min}$ be the earliest observation,
\[
t^\star = t_{\min} + 20~\text{years},
\]
with $\{\text{Date}<t^\star\}$ assigned to training and $\{\text{Date}\ge t^\star\}$ to testing. If the 20-year mark exceeds the sample, the split defaults to the 80th time quantile. This ensures long estimation windows and strict out-of-sample evaluation. Robustness checks with purged walk-forward are described in the appendix.

The trading environment is a daily rebalancing simulator (\texttt{PortfolioEnv}). The state $S_t\in\mathbb{R}^{W\times N\times F}$ is a rolling window ($W{=}30$ days) of standardized features. Actions are portfolio allocations $w_t\in\Delta^{N+1}$ over cash and assets, constrained to the simplex. The mask $M$ ensures weights on unavailable assets are zeroed and renormalized. Turnover costs of $\kappa=5$\,bps per dollar traded are deducted from realized returns. Rewards are log wealth growth net of costs. Cash is always included (\texttt{include\_cash=True}), enabling dynamic de-risking.

\subsection{Policy Architecture, Training Protocol, and Evaluation}
To isolate optimization effects, the policy architecture is fixed across learning rules. Each asset’s recent feature history is first summarized by a temporal encoder, then aggregated via a \emph{cross-sectional attention} layer with a global token. This allows the policy to condition on inter-asset dependencies such as sectoral co-movement, factor spillovers, and substitution effects. Unless otherwise noted, hidden width is $d=64$. The aggregated representation feeds a \emph{flat Dirichlet} head that outputs concentration parameters on the simplex $\Delta^{N+1}$ spanning cash and tradable assets; weights are sampled and renormalized after masking.

We compare three reinforcement learning algorithms under this common attention-based policy: Proximal Policy Optimization (PPO) with clipped gradients and GAE; Advantage Actor–Critic (A2C), with MSE-trained critic and no clipping; and REINFORCE, using Monte Carlo returns with a learned baseline. As a benchmark, we evaluate an equal-weight buy-and-hold portfolio of all tradable tickers. Fine-tuning starts from pretrained checkpoints and applies additional on-policy updates; outcomes appear in Section~\ref{sec:results}.

Training employs Adam with learning rate $3\times10^{-4}$, gradient clipping at $0.5$, and seed $42$. Each update processes 128 days with six updates per epoch; minibatches of 256 are split into micro-batches of 32 for PPO/A2C. Discounting and bootstrapping are $\gamma=0.99$, $\lambda=0.95$ for PPO/A2C; REINFORCE uses Monte Carlo returns with $\gamma=0.99$. Policies rebalance daily under $\kappa=5$\,bps turnover cost, running on GPU when available.

Hyperparameter refinement follows a small-budget grid search over learning rates $\{10^{-4},\,5\!\times\!10^{-5}\}$, epochs $\{3,\,5\}$, and minibatch sizes $\{128,\,256\}$ (micro-batch 32 and updates 6 fixed). Each trial is fine-tuned and evaluated out-of-sample, selecting the best checkpoint per algorithm by Sharpe (alternatives: CAGR or MDD). All runs are seeded and produce reproducible logs and equity curves.

Evaluation is deterministic: stochastic Dirichlet sampling is replaced by its mean. Primary performance is cumulative after-cost wealth
\[
W_T = \prod_{t=1}^T \bigl(1 + r^{\text{port}}_t\bigr),
\]
with $r^{\text{port}}_t$ the net portfolio return. We report CAGR, annualized return and volatility, Sharpe, Sortino, maximum drawdown (MDD), Calmar, hit rate, average gains/losses, skewness, kurtosis, 5\% VaR, 5\% CVaR, tail ratio, and terminal wealth.

\section{Results}\label{sec:results}

\subsection{Baseline Performance}
\begin{figure}[t]
    \centering
    \includegraphics[width=\linewidth]{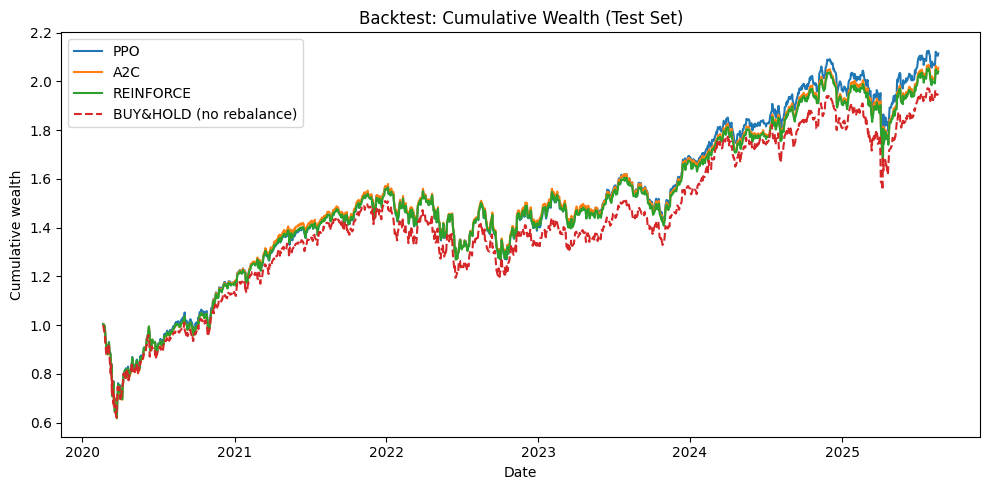}
    \caption{\textbf{Backtest cumulative wealth (2020–2025).} PPO, A2C, and REINFORCE track aggregate market cycles but sustain higher cumulative returns than buy-and-hold, with faster recovery from drawdowns.}
    \label{fig:test_cw}
\end{figure}

Table~\ref{tab:rl_metrics} reports baseline performance, while Figure~\ref{fig:test_cw} plots cumulative wealth over 2020–2025. RL policies consistently outperform buy-and-hold. PPO delivers terminal wealth $2.11$ (CAGR $14.6\%$), followed by A2C ($2.06$, CAGR $14.0\%$) and REINFORCE ($2.04$, CAGR $13.9\%$). Buy-and-hold achieves only $1.94$ (CAGR $12.9\%$). These incremental differences compound into meaningful long-run gains.

Risk-adjusted returns also favor RL. PPO attains Sharpe $0.73$ and Sortino $1.03$, compared with buy-and-hold’s $0.66$ and $0.92$. Calmar is likewise higher (0.39 vs.\ 0.34). The equity curves show that while all methods experienced $\approx38\%$ drawdowns during COVID-19 and mid-2022, RL policies recovered faster and maintained a persistent lead. By 2025, PPO compounded nearly 9\% higher wealth than buy-and-hold. Importantly, RL strategies remain highly correlated with the benchmark, indicating that gains stem from cross-sectional reallocation rather than aggregate market timing. Nonetheless, persistent large drawdowns highlight vulnerability to systemic shocks.

\begin{table}[htbp]
\centering
\caption{Performance comparison of RL policies and buy-and-hold benchmark on the S\&P~500 test set.}
\label{tab:rl_metrics}
\tiny
\begin{tabular}{lcccccccc}
\toprule
Algo & Term. Wealth & CAGR & Ann. Ret. & Ann. Vol. & Sharpe & Sortino & Calmar & MDD \\
\midrule
PPO        & 2.1148 & 0.1462 & 0.1610 & 0.2206 & 0.7298 & 1.0303 & 0.3852 & -0.3796 \\
A2C        & 2.0561 & 0.1404 & 0.1568 & 0.2248 & 0.6975 & 0.9864 & 0.3648 & -0.3848 \\
REINFORCE  & 2.0408 & 0.1388 & 0.1549 & 0.2224 & 0.6963 & 0.9831 & 0.3621 & -0.3833 \\
Buy\&Hold  & 1.9433 & 0.1287 & 0.1460 & 0.2225 & 0.6560 & 0.9207 & 0.3404 & -0.3781 \\
\bottomrule
\end{tabular}
\normalsize
\end{table}

\subsection{Fine-Tuning Results}
We next evaluate fine-tuned policies. Each algorithm was re-trained for six additional updates under different optimizer settings. Table~\ref{tab:ft_best} reports the best configuration per algorithm, while the analysis compares tuned to baseline.

For PPO, fine-tuning materially improved performance. The optimal configuration—learning rate $5\times10^{-5}$, 3 epochs, minibatch 256—achieved Sharpe $0.776$ and CAGR $15.98\%$ with MDD $-37.8\%$. This exceeds the baseline PPO Sharpe $0.730$ and CAGR $14.6\%$, indicating that smaller learning rates and larger batches stabilize updates and enhance exploitation of cross-sectional signals. By contrast, A2C and REINFORCE displayed narrow dispersion across the grid: Sharpe clustered around $0.687$–$0.689$ for A2C and $0.675$–$0.683$ for REINFORCE, with CAGRs $\approx13.7\%$ and $\approx13.5\%$, respectively. Improvements relative to baseline were marginal. This suggests that simpler policy-gradient methods are less responsive to optimizer tuning and inherently constrained by their variance properties.

Taken together, the fine-tuning study shows that PPO is the most adaptable algorithm in this attention-based portfolio setting, with hyperparameter calibration yielding meaningful gains in Sharpe efficiency and long-run growth. A2C and REINFORCE, while competitive with buy-and-hold, offer limited upside from tuning.

\begin{table}[htbp]
\centering
\caption{Best fine-tuned configuration per algorithm (selected by Sharpe on the test set).}
\label{tab:ft_best}
\begin{tabular}{lccccccc}
\toprule
Algo & LR & Epochs & MB & MB$_{\text{micro}}$ & ExtraUpd & Sharpe & CAGR / MDD \\
\midrule
PPO        & $5\times10^{-5}$ & 3 & 256 & 32 & 6 & 0.776 & 15.98\% / $-37.81\%$ \\
A2C        & $5\times10^{-5}$ & 3 & 256 & 32 & 6 & 0.689 & 13.81\% / $-38.40\%$ \\
REINFORCE  & $5\times10^{-5}$ & 3 & 256 & 32 & 6 & 0.683 & 13.57\% / $-38.33\%$ \\
\bottomrule
\end{tabular}
\end{table}

\bigskip

\section{Interpretation \& Economic Mechanisms}\label{sec8}
Our empirical analysis shows that reinforcement learning (RL)–based portfolio policies can deliver incremental yet persistent outperformance relative to a naïve buy-and-hold benchmark. Across PPO, A2C, and REINFORCE, the policies achieve higher long-run growth rates and superior Sharpe and Sortino ratios while remaining exposed to aggregate market cycles. The gains do not stem from market timing, but from systematic cross-sectional reallocation: the policies exploit relative strength, momentum, and volatility-adjusted signals to harvest small inefficiencies that accumulate over time. This mechanism enhances portfolio efficiency without materially deviating from market dynamics.

The central methodological contribution of this study is the integration of an \emph{attention-based policy architecture} into the DRL framework. Prior work has typically relied on convolutional or recurrent encoders to extract temporal features. By contrast, our design combines per-asset temporal encoders with a cross-sectional attention mechanism and a global token. This structure enables the policy to focus dynamically on the most informative assets and features, capturing sectoral spillovers, common factor exposures, and relative-value relationships that conventional architectures treat rigidly. Economically, the attention layer reflects the reality that asset returns are interdependent, and adaptive weighting of these interdependencies improves the extraction of actionable signals from high-dimensional data.

At the same time, the persistence of large drawdowns across all strategies highlights a fundamental limitation. Even with attention, the models remain vulnerable to systemic shocks, indicating that predictive architectures alone cannot fully resolve the inherent risk–return trade-off of equity markets. Future research should therefore incorporate explicit risk-sensitive objectives—such as drawdown constraints, volatility targeting, or utility-based reward shaping—into the attention-enhanced RL framework.

Beyond risk management, several extensions are promising. Enriching the state representation with macroeconomic indicators, sectoral classifications, or alternative data sources could improve recognition of latent regimes and structural shifts. Hybrid approaches that combine attention-based RL with econometric or asset-pricing models may further enhance interpretability and theoretical grounding. Finally, testing robustness across international markets and longer horizons will help determine whether the observed gains reflect structural signal extraction or sample-specific effects.

In sum, this study demonstrates that attention mechanisms can be effectively integrated into DRL portfolio policies, providing both methodological innovation and economic insight. By adaptively weighting cross-sectional dependencies, our framework captures richer interactions than prior DRL models and translates them into higher risk-adjusted returns. While the improvements are incremental, they underscore the value of attention-based architectures as a foundation for the next generation of reinforcement learning applications in finance.

\section{Conclusion}
This study introduces a deep reinforcement learning framework for dynamic portfolio optimization that integrates a Dirichlet policy with cross-sectional attention mechanisms. By enforcing feasibility on the simplex, accommodating tradability masks, and embedding economically motivated frictions such as transaction costs and risk penalties, the proposed design directly connects learning objectives to classical mean–variance trade-offs. The addition of attention layers enables the policy to adaptively weight sectoral co-movements, factor spillovers, and relative-value signals, thereby enhancing the extraction of actionable information from high-dimensional equity data.

Our empirical evaluation on two decades of S\&P 500 data demonstrates that attention-enhanced Dirichlet policies consistently outperform equal-weight and reinforcement learning baselines in terms of terminal wealth and risk-adjusted returns, while maintaining realistic turnover and drawdown profiles. These gains do not reflect aggregate market timing, but rather systematic cross-sectional reallocations that exploit relative strength, momentum, and volatility-adjusted inefficiencies.

The findings contribute to the broader literature on dynamic portfolio choice by showing how reinforcement learning, when grounded in principled parameterization and rigorous economic evaluation, can provide incremental improvements in efficiency and interpretability. At the same time, the persistence of large drawdowns highlights that predictive architectures alone cannot resolve the fundamental risk–return trade-off in equity markets.

Future work should explore extensions that incorporate explicit risk-sensitive objectives such as drawdown constraints or volatility targeting, enrich the state space with macroeconomic or sectoral information, and test robustness across international markets. More broadly, combining reinforcement learning with econometric models may yield strategies that are both more interpretable and more firmly rooted in economic theory.

\section*{Declarations}

The authors did not receive support from any organization for the submitted work. The authors have no relevant financial or non-financial interests to disclose.

\section*{Data Availability Statement}
The data used in this study are publicly available. Financial market data were obtained from \textbf{Yahoo Finance}. Specifically, daily transaction and pricing data for the \textbf{500 constituent stocks of the S\&P 500} were accessed through Yahoo Finance’s open API and data portal, available at:

\noindent\url{https://finance.yahoo.com/}

\end{document}